\documentclass[reprint, 
superscriptaddress,
amsmath,amssymb,aps,
prb
]{revtex4-2}

\usepackage{graphicx}
\usepackage{dcolumn}
\usepackage{bm}
\usepackage{xcolor}
\usepackage{times}
\usepackage[normalem]{ulem}
\usepackage{units}
\usepackage[hypertexnames=false,linktocpage=true,colorlinks=true,linkcolor=blue,anchorcolor=blue,citecolor=blue,filecolor=blue,urlcolor=blue,bookmarksnumbered=true,pdfview=FitB,breaklinks=true]{hyperref}
\usepackage{float}

\begin{document}
\title{Fresnel diffraction imaging of surface nanostructure using coherent resonant X-ray scattering}

\author{L. Burgard}
\affiliation{Department of Physics, University of Wisconsin-Milwaukee, Milwaukee, WI 53201, USA}

\author{C. Neupane}
\affiliation{Department of Physics, University of Wisconsin-Milwaukee, Milwaukee, WI 53201, USA}

\author{A. Balodhi}
\affiliation{Department of Physics, University of Wisconsin-Milwaukee, Milwaukee, WI 53201, USA}

\author{S. Bista}
\affiliation{Department of Physics, University of Wisconsin-Milwaukee, Milwaukee, WI 53201, USA}

\author{S. Butun}
\affiliation{NUANCE Center, Northwestern University, Evanston, IL 60208, USA}

\author{R. Jangid}
\affiliation{National Synchrotron Light Source II, Brookhaven National Laboratory, Upton, NY 11973, USA}

\author{A. Barbour}
\affiliation{National Synchrotron Light Source II, Brookhaven National Laboratory, Upton, NY 11973, USA}

\author{N. Basit}
\affiliation{NUANCE Center, Northwestern University, Evanston, IL 60208, USA}

\author{D. F. Agterberg}
\affiliation{Department of Physics, University of Wisconsin-Milwaukee, Milwaukee, WI 53201, USA}

\author{M. Weinert}
\affiliation{Department of Physics, University of Wisconsin-Milwaukee, Milwaukee, WI 53201, USA}

\author{C. Mazzoli}
\affiliation{National Synchrotron Light Source II, Brookhaven National Laboratory, Upton, NY 11973, USA}

\author{M. G. Kim}
\affiliation{Department of Physics, University of Wisconsin-Milwaukee, Milwaukee, WI 53201, USA}

\date{\today}


\begin{abstract} 

We investigated surface nanostructures on an antiferromagnet MnBi$_2$Te$_4$ using a novel imaging technique, direct (real)-space and real time coherent X-ray imaging (direct-CXI). This technique has provided new insights into antiferromagnetic textures, including the formation of anti-phase antiferromagnetic (AFM) domains and thermal dynamics of AFM domains and domain walls. While this method produces real-space images of AFM textures without requiring a complex imaging retrieval process, its underlying imaging mechanism has not been fully understood, limiting a deep understanding of AFM textures and the information they contain. By investigating the well-defined structural characteristics of the nanostructures fabricated on MnBi$_2$Te$_4$, we elucidate the imaging principle of this novel technique. We find that the observed images can be well explained by Fresnel diffraction integral. Using a simple model from classical optics, our calculations successfully reproduce the experimentally observed images of the nanostructures. This demonstrates that direct-CXI not only provides straightforward real-space imaging but also contains phase information through its Fresnel diffraction integral.
\end{abstract}

\maketitle

\section{introduction}

X-ray imaging techniques have been widely used in various scientific fields. These techniques have been employed to investigate the composition and structural properties of various materials, such as cells in biological research and electronic devices in industrial applications. Some of the most well-known X-ray imaging techniques include X-ray computed tomography (CT),\cite{ct1,ct4,ct2,ct3} dark-field X-ray microscopy,\cite{dark2,dark1,dark3} X-ray topography,\cite{topo1,topo2} X-ray coherent diffractive imaging (CDI),\cite{cdi0,cdi1,cdi2,cdi3} and X-ray ptychography.\cite{cdi1,cdi2,cdi3,pty2,pty0,pty1} Many of these methods are primarily used for studying structural properties of materials. 

In contrast, X-ray imaging techniques suitable for investigating non-structural properties, such as magnetic characteristics, are relatively scarce. Not only are these techniques limited in number, but they also have a narrower range of applicable materials and require further research and development.\cite{cheong1} Examples of such techniques include X-ray photoemission electron microscopy (X-PEEM),\cite{peem1,peem2,peem3} Scanning X-ray nano/microdiffraction,\cite{nanod1,nanod2,nanod3} and X-ray magnetic holography.\cite{holo1,holo2,holo3} Another reason for the lack of X-ray imaging techniques for magnetic properties is that most materials actively studied and utilized in technological applications thus far have been ferromagnets. Since numerous well-established imaging methods, such as Magnetic Force Microscopy\cite{MFM1,MFM2} and Magneto-Optical Microscopy,\cite{magneto1,magneto2} exist for ferromagnetic materials, the development of X-ray-based magnetic imaging techniques has been relatively limited.

Recent studies and reports on methods for easily manipulating antiferromagnetic (AFM) order parameter with electric current have stimulated the emergence of antiferromagnetic spintronics.\cite{spintronics0,spintronics1,spintronics2} Despite such a surge of research activities in AFM spintronics, our understanding of AFM textures (domains and domain boundaries), which are the building blocks of spintronics applications, is not yet complete. This is in part due to the limited experimental methods for studying AFM textures directly. Recently, X-PEEM has been widely used in AFM spintronics research,\cite{xpeem0,xpeem1,xpeem2,xpeem3} but generally only a limited number of techniques exist that can visualize AFM textures.\cite{cheong1}

In 2018, we introduced a novel imaging technique utilizing coherent soft X-ray resonant scattering, which was dubbed X-ray Bragg diffraction Phase-Contrast Microscopy.\cite{mgkim1}  This technique is distinct from X-ray CDI techniques in its ability to generate real-space images in real-time directly from the CCD detector, eliminating the need for mathematically and computationally intensive Fourier transformations. It leverages the principles of the resonant X-ray scattering technique, which is element-specific and sensitive to spin directions within the material, and uses the Bragg diffraction condition.\cite{mgkim1,mgkim2} This technique enables the measurement of the antiferromagnetic order parameter within individual domains. We successfully observed the formation of the anti-phase AFM domains in Fe$_2$Mo$_3$O$_8$\cite{mgkim1} and thermal fluctuations of AFM domains in Ni$_2$MnTeO$_6$.\cite{mgkim2}

Despite the successful demonstrations of this technique, it is still elusive how this technique can generate real-space images while the imaging process occurs at Bragg peaks in reciprocal space. It is even more interesting and puzzling that the real-space images of AFM textures contain interference patterns, especially around the domain boundaries. The lack of our understanding of the imaging mechanism of this new technique gives rise to a challenge in decoding information from the observed AFM textures in the images.

To understand the imaging mechanism of this technique, which we now refer to as the direct(real)-space and real-time coherent soft X-ray imaging technique (direct-CXI), we investigate nanostructures with well-defined physical characteristics on a substrate, rather than AFM domains and domain boundaries, which we have not yet developed a deep understanding. We fabricated various Chromium nano-stripes on an antiferromagnetic material, specifically MnBi$_2$Te$_4$, and imaged them under the resonant (magnetic Bragg) and non-resonant (structural Bragg) conditions. Through the successful imaging of Cr nanostructures, we observe a clear appearance of the interference pattern around the nanostructures in the real-space image and find that the Fresnel diffraction integral can accurately explain our observations. This demonstrates that the direct-CXI not only enables real-space imaging but can also provide information in the reciprocal space, suggesting its potential for phase information retrieval.

\begin{figure}[!t]
    \centering
    \includegraphics[width=1\linewidth]{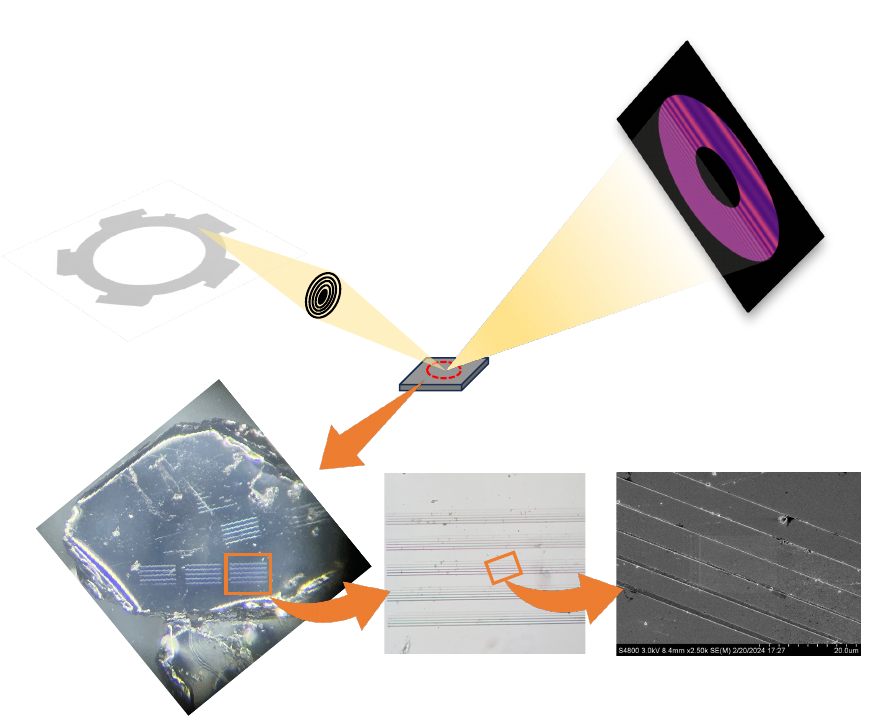}\\
    \caption{Schematic coherent resonant X-ray imaging experiment. (Top) X-rays from the synchrotron pass through the Fresnel zone plate and shine on the sample. The diffracted (reflected) X-rays make images on the two-dimensional CCD detector. (Bottom) Images of the sample. The optical microscope image and SEM image are shown. Various thickness Cr metal stripes (nanostructures) were deposited on a MnBi$_2$Te$_4$ single crystal. } 
    \label{fig1}
\end{figure}

\section{experiment}

MnBi$_2$Te$_4$ single crystals were grown at the University of Wisconsin - Milwaukee (UWM) using the high-temperature solution growth technique as in Ref.~\cite{MBT1}. High-purity elements of Mn, Bi, and Te were used in the ratio of Mn:Bi:Te = 1:10:16 using Bi$_2$Te$_3$ as the flux. The mixture of the elements were placed in an alumina crucible and sealed in a quartz tube in vacuum. Then, the sealed quartz ampoule was placed in a muffle furnace and the temperautre was raised to 900$^\circ C$ with $\approx$ 2$^\circ C$/min and kept at 900$^\circ C$ for 12 hours. The ampoule was then cooled to 595$^\circ C$ for 2 weeks and the ampoule was centrifuged for decanting. We obtained large shiny plate-like single crystals. MnBi$_2$Te$_4$ crystallizes in the trigonal $R\overline{3}m$ and exhibits the AFM transition at $T_N$ = 24 K.\cite{MBT2,MBT3} 

We used the crystal as a substrate for nanostructure fabrication. Thin stripe-shaped nanostructures were fabricated at the Northwestern University Atomic and Nanoscale Characterization Experimental Center. Various thicknesses of Cr stripes (target widths to be 1 $\mu$$m$, 800 $nm$, 500 $nm$, 200 $nm$, 100 $nm$, and 50 $nm$) were deposited on the MnBi$_2$Te$_4$ single crystal (see Figure~\ref{fig1}). The thickness of the Cr stripes was measured using a Scanning Electron Microscope at UWM. The Cr stripes were imaged using a Hitachi S-4800 SEM instrument operated at an accelerating voltage of 3 kV. Their widths are measured using Quartz PCI software and Igor Pro (WaveMetrics). We find that the actual thicknesses of Cr stripes are similar to the target widths of the stripes for 1 $\mu m$, 800 $nm$, and 500 $nm$ stripes. However, Cr stripes thinner than 500 $nm$ were much thicker than the target widths. The actual thicknesses of the Cr stripes are listed in Table.~\ref{tab1}. In this report, for terminological convenience, we will use the target width rather than the actual width when referring to the Cr stripes.

Imaging experiments were performed at the 23-ID-1 CSX beamline at the National Synchrotron Light Source II at Brookhaven National Laboratory. The X-rays were tuned at the Mn $L_3$ edge ($E$ = 640 eV). The sample was mounted at the end of a cold finger of a liquid helium cooled refrigerator on a z-axis diffractometer in a vacuum chamber. The sample was aligned so that the Cr stripes were parallel to the direction of the incoming X-rays. To utilize the resonance enhancement of the magnetic scattering of the sample, measurements were performed at $T$ = 15 K below the antiferromagnetic transition temperature ($T_N$ = 24 K). Snapshot-type images were taken at the antiferromagnetic Bragg peak position (0, 0, 1.5) and in the tail of the structural Bragg peak (0, 0, 3) with a variable exposure time of 0.02 $-$ 0.5 seconds. A Fresnel zone plate (FZP) was used as focusing optics, providing divergent X-rays to the sample, thus leading to the magnification of the obtained images. We can quantify the magnification by measuring the interference pattern. The ratio of the sample-to-detector distance (340 $mm$) and the FZP-to-sample distance (8.4 $mm$) provides a magnification, which is approximately 41. In Fresnel diffraction, the distance between the first and second constructive interference lines is $x_{max} = \sqrt{\frac{3\lambda D}{4}}$, the detector distance $D$ = 340 $mm$, and the X-ray wavelength $\lambda$ = 1.94 $nm$, resulting in $x_{max} \approx$ 22 $\mu m$ for parallel X-rays diffracted from the sample. The observed distance between the first and the second bright lines at the detector is $\approx$ 900 $\mu m$. Thus, the divergence provided by the Fresnel zone plate creates the magnification, $m \approx$ 41 in our experiment setup.The magnification can also be estimated using the beamline parameters. The ratio of the sample-to-detector distance (340 $mm$) and the FZP-to-sample distance (8.4 $mm$) provides a magnification, which is approximately 41.

\begin{table}[t]
    \centering
    \renewcommand{\arraystretch}{1.5}
    \begin{tabular}{|c|c|}
        \hline
        Target width & Measured width \\ \hline
         1 $\mu m$  & 1.005(3) $\mu m$      \\ \hline 
         800 $nm$ & 814.5(8) $nm$   \\ \hline
         500 $nm$ & 509.6(1) $nm$   \\ \hline
         200 $nm$ & 292.9(1) $nm$     \\ \hline
         100 $nm$ & 165.4(8) $nm$     \\ \hline
         50 $nm$ & 135.3(9) $nm$\\ \hline
    \end{tabular}
    \renewcommand{\arraystretch}{1}
    \caption{Measured widths of Cr nanostructures using SEM.}
    \label{tab1}
\end{table}

\begin{figure*}[!t]
    \centering
    \includegraphics[width=1\linewidth]{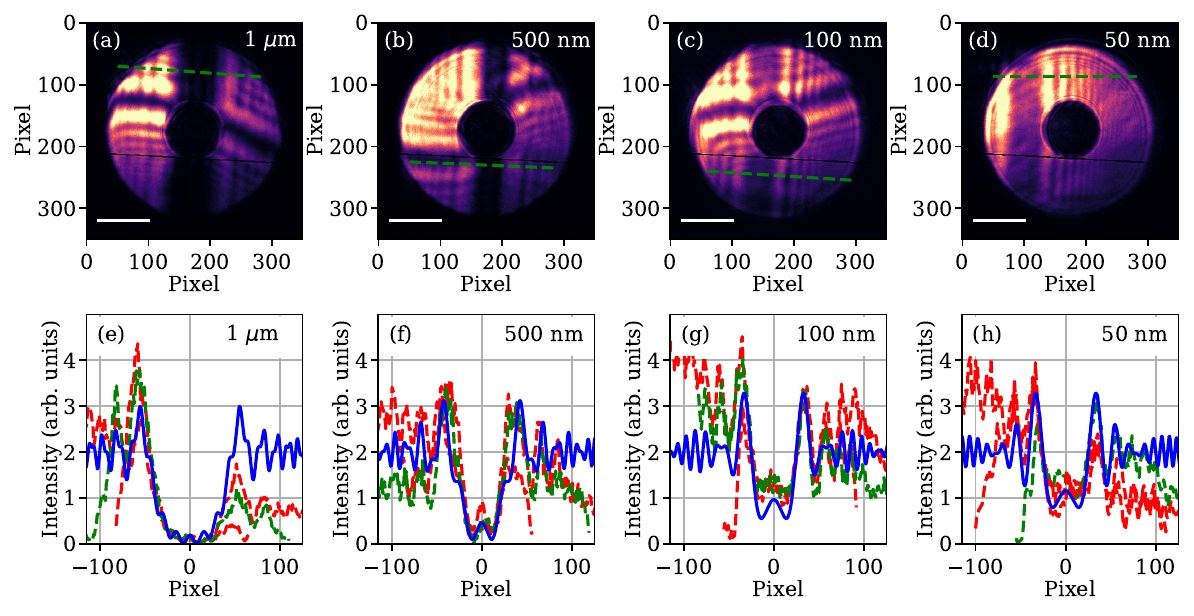}\\
    \caption{(a)-(d) Real space images of Cr stripes of varying thicknesses measured at $Q$ = (0,~0,~1.5) in the resonant diffraction condition. Thick vertical dark straight lines are the image of Cr stripes on the surface (Cr line). Thin parallel lines centered around the Cr line images are interference patterns. Medium thick dark wavy lines are the antiferromagnetic domain boundaries originating from the MnBi$_2$Te$_4$ single crystal. Horizontal thick white bars indicate 2 $\mu m$. A representative one-dimensional (1-D) line cut direction is shown as the green dashed line. It is selected arbitrarily, but normal to the Cr line. (e)-(h) One-dimensional line cuts (red dashed lines) across Cr line images, which are obtained by 2-3 different images of the same Cr line image. A representative line cut is shown as the green dashed line. Blue solid lines are the Fresnel diffraction integral calculations.} 
    \label{fig2}
\end{figure*}

\section{results and discussion}

Figures~\ref{fig2} (a)$-$(d) show direct-CXI images of Cr stripes of various thicknesses (1 $\mu m$, 500 $nm$, 100 $nm$, and 50 $nm$) measured at the AFM Bragg peak position, $Q$ = (0, 0, 1.5) at $T$ = 15 K. The resonant X-ray scattering process at the Mn $L_3$ edge ($E$ = 640 eV) at $Q$ = (0, 0, 1.5) gives rise to the bright intensity in the images. Cr stripes appear as thick vertical dark straight lines in all images, which we refer to Cr line. The horizontal dark wavy lines represent the antiphase AFM domain boundaries originating from the MnBi$_2$Te$_4$ crystal. The formation mechanism of AFM domain boundaries remains under discussion; it is currently attributed to destructive interference arising from phase differences between domains. In this sense, the antiphase AFM domain boundaries are formed by 180$^\circ$ phase differences between the adjacent domains. We observe that Cr lines appear varying in thickness in real space images. The width of the thick dark Cr lines decreases as the Cr stripes become narrower. But, we observe that the width of the Cr lines does not change for Cr stripes narrower than 500 $nm$. We also observe the appearance of an interference pattern as thin vertical dark lines around the Cr lines. 

We plot one-dimensional (1-D) line cuts across Cr lines and show the 1-D cuts in Fig.~\ref{fig2} (e) - (f) with red dashed lines. The two or three distinct red dashed lines represent 1-D line cuts extracted from images measured at slightly different lateral positions of the same Cr line. This is done to capture the interference pattern on both sides of a Cr line due to the small field of view, which is approximately 5 $\mu m$. We observe two features in the 1-D cuts. First, we observe that the intensity at the central position of the Cr line in the image decreases to nearly zero. However, we note that the intensity does not reach zero while exhibiting a distinct structure, such as a wavy (Fig.~\ref{fig2} (e)) or protruding shape (Fig.~\ref{fig2} (f)-(h)). Secondly, outside of the Cr line, the intensity increases and we observe ripples on both sides, which demonstrate the interference patterns. These features indicate that the real space image involves some form of interference phenomenon.

\begin{figure}[!t]
    \centering
    \includegraphics[width=1\linewidth]{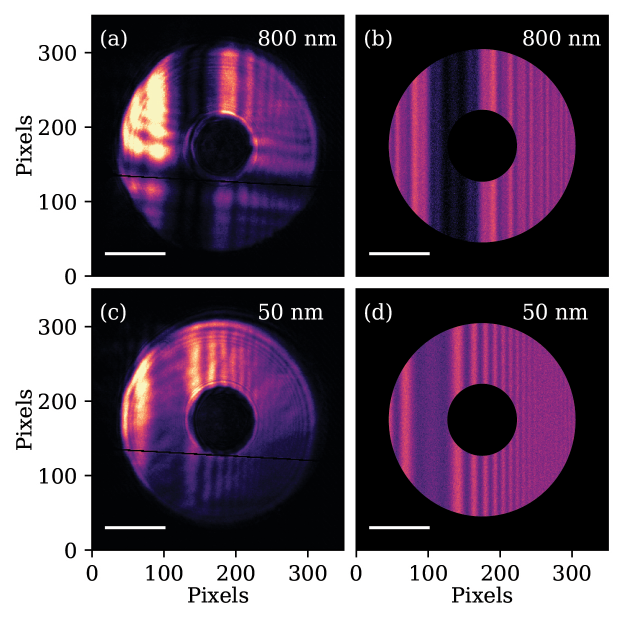}\\
    \caption{Comparison between real space images and Fresnel diffraction integral calculations for the 800 $nm$ stripe, (a) and (b), and for the 50 $nm$ stripe, (c) and (d). Horizontal thick white bars indicate 2 $\mu m$. }
    \label{fig3}
\end{figure}

To understand the observed images of Cr nanostructures, we employ the Fresnel diffraction integral. In our experimental geometry, the length directions of the Cr stripes were placed close to the direction of the incoming X-rays. Such a configuration is a well-known geometry in optics as opaque stripes placed parallel to the beam direction. The Fresnel diffraction integral for the Cr geometry in reflection is given by,\cite{fresnel1,fresnel2,fresnel3}

\begin{equation}
    \begin{aligned}
        I =& \frac{1}{2}\biggl\{[1-C(\nu+\frac{\Delta \nu}{2})+C(\nu-\frac{\Delta \nu}{2})]^2 \\
        &+[1-S(\nu+\frac{\Delta \nu}{2})+S(\nu-\frac{\Delta \nu}{2})]^2\biggr\}.
    \end{aligned}
\end{equation}\label{eq1}


where $C(\nu)$ = \( \int_{0}^{\nu} \cos(\frac{\pi t^2}{2}) \,dt \),  $S(\nu) =$ \( \int_{0}^{\nu} \sin(\frac{\pi t^2}{2}) \,dt \), and $\Delta \nu$ is the width of the thick dark Cr line. Here, $\nu$ is the width of the image of the object in the detector reference frame, which is enlarged in our experiment by the magnification factor, $m \times \nu$. Thus, $m \times ( \nu \pm \Delta \nu/2) = \sqrt{\frac{2 \pi}{\lambda D}}(x+\Delta x/2)$ in the object reference frame where $D$ is the distance between the sample and the detector, $\lambda$ is the wavelength of the X-ray, and $\Delta x$ is the width of Cr stripes. We fit our data using the Fresnel diffraction integral with the least-squares method, using $m$ and $D$ as fitting parameters. The fit results are shown as blue solid lines in Fig.~\ref{fig2} (e)-(h), and the fit values are presented in Table.~\ref{tab2}. Although the fit does not yield an excellent match to the data, we find that it captures the important details of the data; features in the center of the Cr lines and interference patterns on both sides of the Cr lines.  Features such as the wavy and protruding shapes at the center of the Cr lines in Fig.~\ref{fig2} (e)-(h) were successfully reproduced with high accuracy using the Fresnel diffraction integral. The periodicity of the ripples observed on both sides of the Cr lines aligns well with the Fresnel diffraction integral calculations. To highlight this, we present our two-dimensional (2-D) calculations for 800 $nm$- and 50 $nm$-thick Cr stripes in Fig.~\ref{fig3}. Comparisons between Fig.~\ref{fig3} (a) and (b) and between Fig.~\ref{fig3} (c) and (d) show remarkably good matches. It is worth noting that random noise is included in our calculation to mimic the actual data. Despite the good match between the data and the Fresnel integral calculations, we find that the calculation overestimates the widths and the magnification by approximately 1.16 - 1.46 times when we compare the fit values with the actual width of Cr stripes measured by SEM in Table.~\ref{tab2}.

\begin{table}[!t]
    \centering
    \renewcommand{\arraystretch}{1.5}
    \begin{tabular}{|c|c|}
        \hline
                      &  Fit values    \\ \hline
        Magnification &  $\times$ 48    \\ \hline
        1 $\mu m$ & 1.31(1) $\mu m$     \\ \hline 
        800 $nm$ & 1.19(3) $nm$    \\ \hline  
        500 $nm$ & 589(18) $nm$    \\ \hline  
        200 $nm$ & 380(15) $nm$    \\ \hline  
        100 $nm$ & 219(11) $nm$    \\ \hline     
        50 $nm$& 181(16) $nm$   \\ \hline
    \end{tabular}
    \renewcommand{\arraystretch}{1}
    \caption{Fit values using the Fresnel diffraction integral calculations.}
    \label{tab2}
\end{table}

\begin{figure*}[t!]
\centering
\includegraphics[width=1\linewidth]{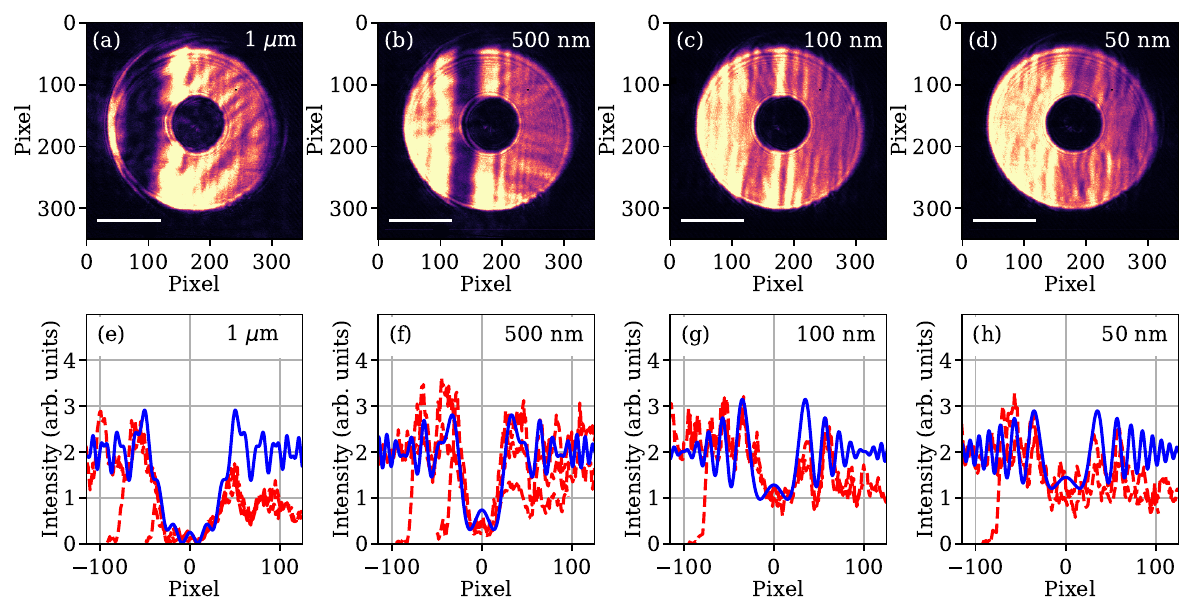}\\
\caption{(a)-(d) Real space images of Cr stripes of varying thicknesses measured in the tail of $Q$ = (0,~0, 3) in the reflection condition. Thick vertical dark straight lines are the image of Cr stripes on the surface. Thin parallel lines centered around the think vertical lines are interference patterns. Horizontal thick white bars indicate 2 $\mu m$. (e)-(h) One-dimensional line cuts (red dashed lines) across Cr lines (thick vertical dark lines), which are obtained 2-3 different images of the same Cr lines. Blue solid lines are the Fresnel diffraction integral calculations.} \label{fig4}
\end{figure*}

\begin{figure}[!]
    \centering
    \includegraphics[width=.9\linewidth]{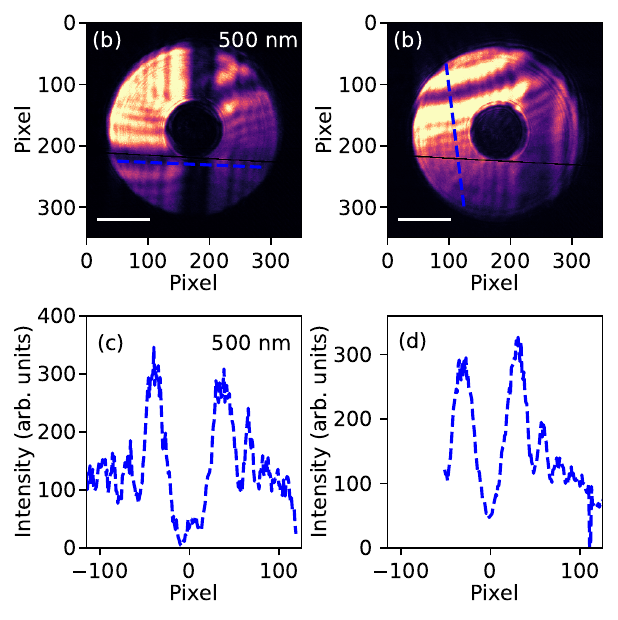}\\
    \caption{Comparison between 500 $nm$ thick Cr stripe and 500 $nm$ AFM domain wall. (a) The 2-D image of the Cr lines of 500 $nm$ (the thick vertical dark line). The 1-D line cut is shown as the blue dashed line. (b) The 2-D image of the AFM domain wall (the thin slanted horizontal dark line). The blue dashed line represents the 1-D line cut direction. (c) The blue dashed line represents the 1-D line cut through the 500 $nm$ Cr line. (d) The blue dashed line is the 1-D line cut through the AFM domain wall of MnBi$_2$Te$_4$.  }
    \label{fig5}
\end{figure}

Figure~\ref{fig4} shows images of Cr nanostructures measured using the intensity from the structural Bragg peak $Q$ = (0, 0, 3). The intensity of (0,~0,~3) is very strong and overwhelms the weak signals of our nanostructures. Thus, we image Cr nanostructures using a fraction of the intensity obtained at the tail of the (0,~0,~3) Bragg peak by rotating the sample slightly away from the Bragg peak position. Figs.~\ref{fig4} (a)-(d) show direct-CXI images of 1 $\mu m$, 500 $nm$, 100 $nm$, and 50 $nm$ thick Cr stripes. These images can be readily compared to Figs.~\ref{fig2} (a)-(d) which were measured at $Q$ = (0, 0, 1.5) at the resonant condition. We observe essentially the same features at the tail of the (0,~0,~3) Bragg peak, which were observed at (0,~0,~1.5) with less clarity in the images. We note that the horizontal wavy lines of the AFM domain boundaries observed in Fig.~\ref{fig2} do not appear in these images (Fig.~\ref{fig4}) because they are in the structural Bragg condition. We present 1-D line cuts across the Cr lines in Fig.~\ref{fig4}. The 1-D line cuts are also obtained from different images of the same corresponding Cr lines to ensure capturing interesting features. At the center of the line cut, instead of observing the intensity decreasing to zero, we observe some structures such as ripples in Fig.~\ref{fig4} (e) and humps in Fig.~\ref{fig4} (f)-(h). Although such structures in these images appear noisier than those observed in Fig.~\ref{fig2} (e)-(h), the 1-D line cuts of the Cr lines capture the fundamentally identical features that appear in Fig.~\ref{fig2}. 

The successful application of the Fresnel integral is possible because our imaging falls within the Fresnel diffraction limit. The Fresnel diffraction limit can be determined by evaluating the Fresnel number, $\mathrm{N}_\mathrm{F} = a^2/L\lambda$,\cite{fresnel4} where $a$ is the characteristic length of the object, which corresponds to the thickness of Cr stripes, $L$ is the distance between the aperture and the detector which is approximately $D$, and $\lambda$ is the wavelength of the X-rays. If $\mathrm{N}_\mathrm{F} > 1$, the system is within the Fresnel diffraction regime, whereas if $\mathrm{N}_\mathrm{F} \ll 1$, it falls within the Fraunhofer diffraction limit. In our experiment, the observed thicknesses of the Cr stripes on the detector are approximately 48 times larger than the values calculated from the Fresnel diffraction integral, which was determined based on the given experimental parameters and the thicknesses of the Cr stripes on the sample. This magnification is attributed to the FZP. However, in the absence of the FZP in the Fresnel integral calculation, achieving the same result would require the Cr stripe to be approximately 48 times thicker. Therefore, it is necessary to apply the magnification factor to the characteristic length, $m \times a$. We find that the Fresnel number becomes 1 when the Cr stripe thickness is approximately 550 $nm$. Therefore, our image is in the Fresnel limit for Cr stripes thicker than approximately 550 $nm$. This result explains why the thickness of the dark line in the Cr line image varies between 1 $\mu m$ and 500 $nm$, but remains unchanged below 500 $nm$.

Lastly, we comment on the difference between the AFM domain walls and the Fresnel diffraction integral on the surface structure (Cr stripes). Figure~\ref{fig5} shows 1-D line cuts of the AFM antiphase domain walls in MnBi$_2$Te$_4$ crystal at $T$ = 15 K and the 500 $nm$ Cr stripe. We compare the AFM domain wall to the 500 $nm$ Cr line because the Magnetic Force Microscopy measurements showed that the thickness of the domain wall is $\approx$ 500 $nm$.\cite{MBT4} We can clearly see that the 1-D line cut of the AFM domain wall are very different from that of the surface structure. However, we find that the interference pattern appears around the AFM domain wall (Fig.~\ref{fig5} (b)). This implies the importance of Fresnel diffraction in imaging AFM domain walls. Further theoretical study is necessary to understand the mechanism of imaging AFM domain walls in the Fresnel diffraction integral.

\section{conclusion}

We successfully imaged Cr nanostripes using the direct-CXI technique. In this experiment, we employed two different imaging approaches: first, by utilizing the AFM resonant signals from the MnBi$_2$Te$_4$ single crystal on which the Cr stripes were deposited, and second, by using the structural Bragg peak intensity of the material. While the two imaging approaches exhibited differences in clarity and signal-to-noise ratio, they consistently revealed the same essential features, including a distinct pattern at the center of the Cr line images and observable interference patterns around them. The observed images were well described by the Fresnel diffraction integral. For geometrical objects, this technique is in the Fresnel limit for the size larger than approximately 500 $nm$. The direct-CXI technique is expected to serve as a powerful tool for imaging AFM domains and domain walls, with broad applicability in the field. Despite the fundamental difference between a geometrical object and AFM domain boundaries, our findings contribute to a better understanding of the underlying measurement principles of this technique, further enhancing its potential for future studies.

\begin{acknowledgments}
This work was supported by the University of Wisconsin-Milwaukee. DFA and MW were supported by National Science Foundation Grant No. DMREF 2323857. This work made use of the NUFAB facility (RRID:SCR\textunderscore017779) of Northwestern University’s NUANCE Center, which has received support from the SHyNE Resource (NSF ECCS-2025633), the IIN, and Northwestern’s MRSEC program (NSF DMR-2308691). This research used resources at the 23-ID-1 beamline of the National Synchrotron Light Source II, a DOE Office of Science User Facility operated for the DOE Office of Science by Brookhaven National Laboratory under Contract No. DE-SC0012704. RJ, AB, and CM also acknowledge the resources made available through BNL/LDRD\#19-013.

\end{acknowledgments}

\section{references}

\bibliography{Fresnel}

\end{document}